# MVA Transfer Pricing

Wujiang Lou[1]
Oct 12, 2015. Updated July 28, 2016.

**Abstract**

This article prices OTC derivatives with either an exogenously determined initial margin profile or endogenously approximated initial margin. In the former case, margin valuation adjustment (MVA) is defined as the liability-side discounted expected margin profile, while in the latter, an extended partial differential equation is derived and solved for an all-in fair value, decomposable into coherent CVA, FVA and MVA. For uncollateralized customer trades, MVA can be transferred to the customer via an extension of the liability-side pricing theory. For BCBS-IOSCO covered OTC derivatives, a market maker has to charge financial counterparties a bid-ask spread to transfer its funding cost. An IM multiplier is applied to calibrate to external IM models to allow portfolio incremental pricing. In particular, a link to ISDA SIMM for equity, commodity and fx risks is established through the PDE with its vega and curvature IM components captured fully. Numerical examples are given for swaps and equity portfolios and offer a plausible attribution of recent CME-LCH basis spread widening to elevated MVA accompanying dealers' hedging of customer flows.

**Keywords:** initial margin, margin valuation adjustment (MVA), ISDA SIMM, liability-side pricing, coherent CVA and FVA, CME-LCH basis.

## 1. Introduction

As September 2016 rollout of margin requirements for non-centrally cleared OTC derivatives (BCBS-IOSCO 2015) is fast approaching, derivatives pricing with initial margin (IM) remains a significant challenge as IM, essentially a capital measure of potential future exposure based on historical data, forces its way into a pricing model under the risk neutral measure. At a fixed future time, for example, the short rate would evolve and accumulate filtration under both the risk neutral and the physical measure. As such, a model implementing initial margin strictly to its CCP or BCBS-IOSCO specifications would have to resort to Monte Carlo simulation. To be truthful to specs, a brute force simulation would simulate a path in the pricing measure, adopt it as an extended historical set of data in the physical measure, update the scenarios including stress scenarios, and conduct a separate or layered simulation in the physical measure for initial margin, conditional on the path. IM's path dependency destroys the Markovian property of the price process afforded by existing rate models and limits our modeling and analytics options.

---

[1] The views and opinions expressed herein are the views and opinions of the author, and do not reflect those of his employer and any of its affiliates. The author would like to thank Naosuke Nakamura, Marco Ossanna, Gary Li and an anonymous reviewer for valuable suggestions. Earlier versions are titled 'Initial margin funding cost transfer pricing and MVA'.

Green and Kenyon (2015) calculate initial margin as Value-at-Risk (VaR) by simulation using a fixed set of exogenously determined historical scenarios, ignoring scenario updating, which is questionable for long duration swap portfolios as BCBS stipulates a historical period of not exceeding 5 years. Brigo and Pallavicini (2014) define initial margin in the pricing measure, as the q-quantile of a normal variate with its variance matching that of the conditional close-out amount in the margin period of risk. While forward scenario updating is captured, connection to the physical measure is lost. On the other hand, IM calculations have largely been standardized, for example, ISDA's standard initial margin model (SIMM), and made available from vendors, providing an opportunity for models to focus on the pricing aspect.

Technicality aside, a derivatives desk's concern is how to manage or transfer the added cost of posting initial margin. When two CCP clearing members trade CCP cleared derivatives or two covered counterparties[2] trade non-centrally cleared OTC derivatives, both parties post initial margin and both incur funding costs. The bilateral trade *alone* does not offer any economic way of cost transfer, unless one party takes the role of a market maker who then packs the cost into his bid-ask spread. A more meaningful case of cost transfer occurs when a clearing member trades with a non-clearing client and has to fund initial margin on its hedge trade through the CCP, or similarly when a covered entity trades with a non-covered one and hedges with another covered entity. In such a case, the client side pricing effectuates the cost transfer.

This paper contributes a balanced approach by computing IM as a delta-approximated VaR in the pricing measure that naturally captures forward and local updating, with a portfolio specific multiplier allowing calibration to external models meeting CCP or BCBS-IOSCO requirements. This approach is advantageous in that it preserves usual Markov properties and allows PDE and its efficient solutions, avoiding the need of exclusively relying on a layered local simulation that is computationally demanding and often subject to dangerous shortcuts. In particular, ISDA's SIMM can be fully integrated into the PDE for equity, commodity and fx derivatives. IM funding cost is incorporated in the back-to-back derivatives pricing setting in the liability-side pricing framework where CVA and FVA are coherently defined to abide by the law of one price, different from Green and Kenyon (2015) which is derived from Burgard and Kjaer (2011) where funding cost is considered from a private value perspective.

This paper is structured as follows. Section 2 incorporates a redundant initial margin account into the back-to-back swap hedging economy to extend the liability-side Black-Scholes-Merton partial differential equation and defines MVA for an exogenously prescribed initial margin profile. Section 3 employs the delta approximation of VaR to compute initial margin and derives a more analytically tractable PDE for the fair value with initial margin funding costs. Section 4 discusses BCBS-IOSCO covered OTC derivatives, incorporates gamma and vega approximation to link with ISDA's SIMM, and shows that the cost transfer has to be in the form of a bid/ask spread. Section 5 applies finite difference and Monte Carlo simulation to compute standalone single trade and swap and equity option portfolios MVA. Section 6 concludes.

---

[2] Per BCBS-IOSCO 2015, all financial firms and systemically important non-financial entities ("covered entities" or financial counterparties) engaging in non-central cleared OTC derivatives are required to post full variation margin and bi-way segregated initial margin, subject to a minimum level of activity.

## 2. Liability-side PDE of Uncollateralized Trades

Initial margin is essentially a reserve or defaulter-pay capital for potential future exposure (PFE) during a margin period of risk (MPR). For capturing funding cost purposes, we idealize this reserve as redundant[3].

### 2.1. Redundant initial margin account

Consider a hypothetical bank (party B, a CCP clearing member) clears a derivatives portfolio and posts IM to a CCP. If the bank defaults, the bank's trade with CCP would get assigned to an operating clearing member. In an orderly and timely transfer, derivatives portfolio would be taken over at their fair value which is fully margined, so that B's initial margin with the CCP would get returned in whole. If party B's client defaults, the client side trades settle and the CCP side swaps can be terminated by B immediately with a full return of incremental IM. In an efficient default settlement that takes no time to occur, initial margin is thus redundant.

Liquidation of a defaulting CCP member of course does not happen instantaneously. A default and subsequent margin settlement window could take days and market could move negatively during such a margin period of risk (MPR), resulting in losses. In the standard credit risk pricing approach, the expected loss (EL) is priced in while the unexpected loss (UL) is treated as capital. Because the period is only couple days, the loss distribution's EL is very small and its UL far outweighs EL. Initial margin basically is a VAR-kind of UL measure, a reserve rather than an economic loss, and is expected to be fully returned. While the probability of exhausting the IM is only 1%, the probability of some partial yet significant loss (say 30%) to IM is not necessarily small.

Green and Kenyon (2015) also separate potential losses from IM, arguing that any loss shall be considered from B's capital account, which would command a capital charge to the client to compensate the propensity of loss. In fact, it goes beyond being redundant as IM is used to fund the derivatives. Obviously, funding IM requirement rather than coping with potential loss to its IM when the firm defaults is business for a going-concern, providing further support for a bank to adopt such a stylized redundant initial margin account to evaluate valuation impact and dictate price transfer. In formalizing IM as a redundant reserve, we agree with Green and Kenyon to treat loss to IM separately[4].

Meanwhile, the cost of maintaining IM accumulates over the derivatives' lifespan in years and is far greater than cost of replenishing potential IM loss over the MPR in days. Therefore a redundant IM account is a reasonable assumption to make in the presence of a short MPR, to evaluate its valuation impact and dictate price transfer.

### 2.2. PDE with initial margin funding cost

Suppose an uncollateralized customer (party C) enters into an interest rate swap (swap #1) with party B. The bank hedges the interest rate risk by continuously trading in a number of CCP swaps with a dealer (party D), another CCP member firm. Swap #1 has a unit notional. The CCP swap hedges have dynamic notional of $\Delta_k$, $k=1, 2, ..., K$. CCP swaps are fully collateralized in cash and priced at the risk-free rate $r(t)$. Let $V_k^*$ denote the fair value of k-th CCP swap per unit notional, $L_s$ the variation margin account balance under the CCP swap clearing agreement, then $L_s = \sum \Delta_k V_k^*$, assuming that collaterals are maintained perfectly and continuously. If the client swap is a plain swap directly cleared in CCP, then the hedge can be done in a back-to-back fashion (Lou 2016), i.e., $K=1$.

---

[3] Redundant reserve is a common term in insurance literature, defined as the surplus of the statutory reserve over the economic reserve.

[4] An effort has been made since to capture loss to IM, see "Capital pricing during margin periods of risk and repo KVA", Wujiang Lou, available in SSRN.

Furthermore let $L_I \geq 0$ be an adapted process denoting the segregated IM amount posted by B to CCP on the hedge portfolio. The collateral posted to the account is assumed to be cash and receives $r$ on its balance.

On the client side, there is neither variation margin nor initial margin between party B and C. Additionally, party B maintains a bank account and short term borrowing account. Let $M_t \geq 0$ be the bank account balance that earns interest at $r$, $N_t \geq 0$ the borrowing account balance that pays par rate $r_N(t)$, $r_N(t) \geq r(t)$.

For modeling purposes, a voluntary cash deposit account $L_t$ is added to the economy to record the mutually funded cash deposit under the liability-side pricing principle (Lou 2015). The main idea is that the client is indifferent to making a cash deposit or loan to the bank so long as the loan earns its current market debt rate as the money loaned could be raised from the debt market. Write $L_t = L_t^+ - L_t^-$, $L_t^+$ the cash amount loaned by party C to B that pays C's cash debt interest rate $r_c(t)$, and $L_t^-$ the cash loan by B to C earning B's interest rate $r_b(t)$. Furthermore, the fair value of swap #1 is fully covered by the deposit, i.e., $L_t = V_t$.

The wealth equation of the hedged swap economy reduces to the balance of the bank account and the debt account,

$$\pi_t = M_t + (1-\Gamma_t)(V_t - L_t - N_t - \sum \Delta_k V_k^* + L_s + L_I) = M_t - (1-\Gamma_t)(N_t - L_I),$$

where $1-\Gamma$ is party B and C's joint survival indicator. To fully replicate the pre-default portfolio, we set $\pi = 0$. Consequently $M_t = 0$ and $N_t = L_I$, reflecting the obvious that, under this setup, the initial margin amount is the only item left to be funded.

Now if the bank defaults, the cash loan as a receivable to C and the derivative as a payable to C set off under the ISDA set-off provision and there is no default settlement, avoiding the need of an auction style liquidation process which would entail a margin period of risk. The bank's trade with CCP gets assigned to another operating clearing member with zero loss to IM. If party C defaults, the client side trade again settles trivially under set-off and CCP swaps can be terminated by B immediately with a full return of IM.

The returned IM can be used to pay back the borrowing amount $N_t$, so that there is no gain or loss to both issuer and holders of $N_t$. In this idealized setting, $N_t$ has an endogenous par recovery and is credit risk free. In fact, $N$ can be seen as a secured funding note with IM as its collateral[5].

Excluding the finite number of discrete swap payment dates, there is no swap cashflow pre-default, so the financing equation is written as follows,

---

[5] Bank's claim, or residual interest, on IM posted to the CCP or BCBS-IOSCO covered counterparties can be assigned to a trust which then issues certificates to finance the IM. The certificates could be made recourse on the bank, in a form similar to a credit linked notes. The coupon rate of the certificates reflect market's credit risk assessment, i.e., propensity of IM losses.

$$\sum d\Delta_k (V^*_k + dV^*_k) + dL - r_c L^+ dt + r_b L^- dt - dL_s + rL_s dt - dL_I + rL_I dt + dN - r_N N dt = 0$$

Plugging in collateral account balances, this becomes,

$$dV - r_c V^+ dt + r_b V^- dt - (r_N - r)L_I dt - \sum \Delta_k (dV_k^* - rV_k^* dt) = 0$$

Let $\rho$ be the interest rate governing swap payments, driven by multiple factors $x_j$, $j = 1, 2, ., J$, under a proper risk neutral measure Q, where each factor is modeled as a diffusion process, $dx_j = a_j dt + b_j dW_j$ with $W_j$ being independent Brownian motions.

$V^*$ then solves the following PDE,

$$(\frac{\partial}{\partial t} + A_t)V_k^* - rV_k^* = 0.$$

where $A_t$ is the generator of the vector diffusion process, $A_t V = \sum_j a_j \frac{\partial V}{\partial x_j} + \frac{1}{2}\sum_j b_j^2 \frac{\partial^2 V}{\partial^2 x_j}$.

. Suppose that all short rates are a function of $\rho$. Noting the above and applying Ito's lemma to $V_t$ and $V^*$ lead to

$$(\frac{\partial V}{\partial t} + A_t V + r_b V^- - r_c V^+ - (r_N - r)L_I)dt + \sum_j b_j \frac{\partial V}{\partial x_j} dW_j - \sum_k \Delta_k \sum_j b_j \frac{\partial V_k^*}{\partial x_j} dW_j = 0$$

Assume that a unique swap delta hedge exists, $\frac{\partial V}{\partial x_j} = \sum_k \Delta_k \frac{\partial V_k^*}{\partial x_j}$, and set $dt$ term to zero, we arrive at,

$$(\frac{\partial}{\partial t} + A_t)V - r_e V - s_I L_I = 0$$

where $r_e(t) = r_b(t)I(V(t)<0) + r_c(t)I(V(t) \geq 0)$ is the switching discount rate and $s_I = r_N - r$ is the funding rate of IM. On swap payment dates, $V$ jumps in accordance with the jump in the swap cumulative dividend process $D(t)$. Applying Feynman-Kac theorem to arrive at the pricing formulae,

$$V(t) = E_t^Q[\int_t^T e^{-\int_t^s r_e du}(dD_s - s_I L_I(s)ds)]$$

When IM is redundant, the uncollateralized swap can be replicated by CCP swaps as before, in both risk sense and cashflow sense. The latter can be seen from $\Delta_T = 1$, necessarily from the terminal boundary condition $V_T = V^*_T = D_T$, where $D_T$ is payoff of the swap at maturity $T$.

Let $U = V^* - V$ be the valuation adjustment to the risk-free price $V^*$, $U$ is given by

$$U(t) = CRA + MVA,$$
$$CRA = E^Q_t [\int_t^T (r_e - r) V^*(s) e^{-\int_t^s r_e du} ds],$$
$$MVA = E^Q_t [\int_t^T s_I L_I(s) e^{-\int_t^s r_e du} ds]$$

In addition to the counterparty risk adjustment (CRA), the funding cost of initial margin introduces a new adjustment -- margin valuation adjustment (MVA). If initial margin $L_I$ is known or computed exogenously, for instance, by means of referencing a standardized schedule, the PDE can be solved by finite difference methods, avoiding the need of running Monte Carlo simulation. Decomposition of CRA into bilateral, coherent CVA and FVA can be done by solving the PDE with shifts in each parties' synthetic or cash curves (Lou 2015) and MVA is obtained by subtracting CRA from the total valuation adjustment. Because MVA itself is part of the fair value, strictly speaking, the formulae for MVA is recursive. An implementation directly calculating MVA however risks to have open IR01 that could attract capital in the same way as FVA when defined imprecisely (Lou 2016), while a PDE based solution is not concerned.

The MVA formulae is different from Green and Kenyon (2015) where the discount factor is the product of the risk-free discount factor and the joint survival probability and the IM funding rate is tied to the dealer's unsecured rate $r_b$, a result of assuming the risk-free close-out and extending Burgard and Kjaer (2011), which innovatively and yet controversially introduces a bank's own funding cost into derivatives fair value. Our formulae comes from the liability-side pricing theory which deviates away from Burgard and Kjaer, and has such desirable aspects as law-of-one-price conforming, FVA relating to bond-CDS basis or liquidity basis, total counterparty risk adjustment (CRA=CVA+FVA) of a bond consistent with bond pricing, and explicit rule for which discount curve to use[6].

Also the IM funding rate is tied to the dealer's unsecured rate $r_b$ in Green and Kenyon (2015), while we have left $r_N$ open. Setting $r_N$ to $r_b$ is of course an option, although a conservative and expensive one, as IM is certainly of collateral value, due to its tail loss nature.

## 3. Initial Margin Valuation Adjustment with Delta Approximation

To gain better insight of the impact of funding cost of initial margin on derivatives pricing and its contribution relative to other factors such as counterparty credit risk and funding measure, we adopt delta approximation of VaR. Consider a scenario shift $\delta x_j$ associated with IM calculation, to the first order the change in the CCP swap portfolio value can be written,

$$\sum_k \Delta_k \delta V_k^* = \sum_k \Delta_k \sum_j \delta x_j \frac{\partial V_k^*}{\partial x_j} = \sum_j \delta x_j \frac{\partial V}{\partial x_j}$$

---

[6] In "Unlocking the FVA Debate" (W Lou, 2015, available in ssrn), Lou considers three segments of the OTC derivatives market, the competitive market, the non-competitive market, and the limited competitive market, and proposes to apply law-of-one-price model to competitive market, private value model (Burgard and Kjaer and its variants) to non-competitive market.

In particular, for a one-factor diffusion model, $d\rho = adt + bdW$, IM can be estimated as

$$L_I = \left|\frac{\partial V}{\partial \rho}\right| \alpha b \sqrt{\delta}$$

where α is the normal deviate corresponding to a given confidence interval, e.g., 2.33 for 99% one-sided quantile, and δ is the length of the VaR period in years[7].

Plugging into the PDE to obtain,

$$\frac{\partial V}{\partial t} + (a - s_I \alpha b \sqrt{\delta} sign(\frac{\partial V}{\partial \rho})) \frac{\partial V}{\partial \rho} + \tfrac{1}{2} b^2 \frac{\partial^2 V}{\partial \rho^2} - r_e V = 0$$

IM funding cost now shows in the PDE's delta term. Intuitively all capital measure involves a position's price variability which for a derivative is governed by its delta, to the first order.

To gain some insight, let's consider the extra counterparty risk protection afforded by initial margin that can be alternatively accomplished by an overcollateralized variation margin account. In fact, IM resembles the collateral posted up to a security lender in a short stock trade where the additional cash collateral is determined by haircut. Because haircut is a percentage applied to fair value, it would affect the rate of return on *V* but pricing can proceed in the usual Markovian way. For derivatives, however, we cannot set a fixed haircut exactly because of its delta. A deep out-of-the-money bond option nearing expiry for example has zero chance of loss, therefore a minimal or zero haircut suffices. A deep in-the-money option on the same bond has the same risk profile as holding a unit of bond, so its haircut will follow that of the underlying bond, which is captured as $\alpha b \sqrt{\delta} = \alpha \sigma P \sqrt{\delta}$ where P is bond price and σ is bond price volatility. Approximation IM with $\left|\partial V / \partial P\right| \alpha b \sqrt{\delta}$ is simply multiplying the underlying bond haircut $\alpha b \sqrt{\delta}$ with the delta, exactly reflecting the derivative's leverage effect by taking the underlying's haircut timed delta as the derivative's value haircut. In this sense, the derivative's haircut is dynamic and when it enters the PDE, it is intuitively captured in the delta term.

---

[7] BCBS has 10 business days, CME is 5 day VaR with 99%, and LCH 99.5% ES 5 days for members or 7 days for non-member clients.

## 3.1. Swap MVA

For a payer swap, its DV01 is always positive and for a receiver swap negative. The sign function in the delta term is resolved and the PDE returns to its familiar liability-side pricing form, for example, for a receiver swap,

$$\frac{\partial V}{\partial t} + (a + \lambda b)\frac{\partial V}{\partial \rho} + \tfrac{1}{2}b^2 \frac{\partial^2 V}{\partial \rho^2} - r_e V = 0,$$
$$\lambda = s_I \alpha \sqrt{\delta}.$$

The swap PDE's delta term now sees a drift in the amount of $\lambda b$, suggesting another equivalent Martingale measure P~ for computing the IM adjusted fair value. This delta approximation induced measure however should not be confused with the real world measure P and $\lambda = s_I \alpha \sqrt{\delta} > 0$ is not the market price of risk. The increased drift in the short rate corresponds to lowered price for zero coupon bonds, meaning less pv on the received fixed rate payments, thus a lower swap npv. For an ATM swap (at par rate) priced at zero npv in the pricing measure Q, its npv in P~ would be negative. To bring it back to zero, the fixed rate has to be increased so that the ask or receiver swap rate is higher than the par rate.

For a payer swap, $\lambda = -s_I \alpha \sqrt{\delta}$. This lowered short rate corresponds to a higher zero price, making the payer ATM swap again negative. To bring it back to par, the swap rate has to be reduced below the par rate. This exactly amounts to a lowered bid. Now we see that the funding cost of initial margin contributes to a swap rate bid/ask spread.

For vanilla swaps, the uncollateralized fair value V can now be written as a risky discounted expectation under a changed measure P~ where the rate process has a shifted drift,

$$V^{P\sim}(t) = E_t^{P\sim}[\int_t^T e^{-\int_t^s r_e du} dD_s].$$

The fair value without initial margin, denoted as $V^Q$, is simply the same expectation under the risk-neutral measure Q. MVA is the difference between these two fair values, MVA = $V^Q$ - $V^{P\sim}$.

Using delta approximation to compute VaR and IM is not new, but introducing it into a PDE setting allows us to apply traditional numerical techniques such as a tree or lattice model when the dimension is low such is the case at a single trade level or a granular portfolio, e.g., a commodity portfolio of a single underlying or rate. For standalone vanilla swaps, the Crank-Nicholson FD scheme (Lou 2016) can be used directly with an adjusted drift coefficient. The regression/simulation procedure designed to handle the discount switch can be modified to handle the delta switch as well, allowing computing MVA alongside CVA and FVA.

## 3.2. IM multiplier

While VaR is measured in the real world, risk metrics involving future such as potential future exposures (PFE) are debatably measured in the real world, or the risk neutral world, or both (Stein 2015), with noticeable differences. Initial margin, essentially a VaR measure of OTC derivatives portfolio' PFE, is subject to both. Its delta approximation, while allowing us to keep the pricing task in the Markovian domain and capturing IM's forward profile, is solely under the risk neutral measure and does not touch upon the non-procyclicality requirement, which is met with historical simulations in the real world. To accommodate each CCP's specific initial margin calculation method, a constant multiplier is incorporated into parameter $\alpha$.

Specifically, rewrite $\alpha = \alpha_q \eta$, where $\alpha_q$ is the q-quantile and $\eta$ is a multiplier specific to the Chicago Mercantile Exchange (CME), for instance. Conceptually this multiplier also captures the requirement of a historical stress period where the short rate volatility has increased. Its calibration to CCP's IM methodology and choices of historical scenarios re-establishes MVA's link to the real world.

A dealer's implementation can calibrate the multiplier to its whole portfolio with a CCP and apply it to a new trade on an incremental basis. When adding a new trade to a dv01 balanced swap portfolio, for example, the dealer may not charge the full incremental cost of financing IM to the new trade, on the ground that bid/ask spread collected on trades in the portfolio could shoulder a large portion of the cost. Correspondingly the multiplier $\eta$ could be made smaller to reflect this reduced charge. This leads to an incremental pricing scheme, illustrated later with an explanation of CME-LCH basis spread.

A firm developing internal quantitative IM models can further factor the multiplier into a stress volatility multiplier and a portfolio netting multiplier. The latter can be taken as 0.4+0.6*NGR deduced from BCBS-IOSCO's standardized IM method where the net standardized initial margin equals to 0.4+0.6*NGR times aggregated products of the standardized margin rate and the gross notional size of each derivatives contract. NGR is defined as the net replacement cost over the gross replacement cost[8] for transactions subject to legally enforceable netting agreements.

With IM calculation complimented with a multiplier, IM funding cost now rests on the funding rate charged on IM. As discussed earlier, IM bears little credit risk and could be used to secure an issuance of funding notes $N$. If $N$ at the same time has a recourse to the general credit of the bank in a senior unsecured rank[9], the eventual loss given default on $N$ is very small. We assume that LGD applicable to $N$ is zero, so that the PDE derived stands as is, while the IM funding rate could incorporate the firm's senior unsecured funding rate or even capital charge rate with the effect of leverage ratio incorporated as shown in Section 5.

For incremental pricing, as a firm could benefit from reduced IM and its funding cost, the multiplier can be made algebraic with a positive value indicating the existing portfolio is long delta, negative short delta. Furthermore, one can adopt an asymmetric multiplier scheme where separate multipliers ($\eta^+$ and $\eta^-$) are applied to the positive delta and negative delta locally. For uncollateralized options, similar PDE can be obtained, either by repeating the CCP replicating exercise in Section 2 or by conducting usual dynamic hedging with underlying stocks (Lou 2015) with an initial margin account added. To highlight MVA's contribution to option bid/ask, we show below the bid and ask side fair values $V_b$ and $V_a$'s PDEs.

---

[8] NGR made out of net dv01 vs gross dv01 is more appropriate given IM's link to delta.
[9] The debt can be issued for instance as a structured note of the bank or swap dealer that references the loss to IM.

$$\frac{\partial V_b}{\partial t} + rS\frac{\partial V_b}{\partial S} - s_I \alpha_q \sigma \sqrt{\delta} S(\eta_+ (\frac{\partial V_b}{\partial S})^+ + \eta_- (\frac{\partial V_b}{\partial S})^-) + \tfrac{1}{2}\sigma^2 S^2 \frac{\partial^2 V_b}{\partial S^2} + r_b V_b^- - r_c V_b^+ = 0,$$

$$\frac{\partial V_a}{\partial t} + rS\frac{\partial V_a}{\partial S} + s_I \alpha_q \sigma \sqrt{\delta} S(\eta_+ (\frac{\partial V_a}{\partial S})^+ + \eta_- (\frac{\partial V_a}{\partial S})^-) + \tfrac{1}{2}\sigma^2 S^2 \frac{\partial^2 V_a}{\partial S^2} - r_b V_a^+ + r_c V_a^- = 0$$

We see that on the bid side, the margin funding cost term (the third term) deducts value from bid while on the ask side (second PDE), it adds value to ask-side, thus creating a wider bid/ask spread. The delta approximation, though crude at the first sight, does capture the shortened duration effect of a swap or options portfolio because it is applied locally. It could be complemented with gamma and vega approximation as shown in the next section.

## 4. BCBS-IOSCO Non-centrally Cleared OTC Derivatives

BCBS-IOSCO covered trades are no longer uncollateralized as in Section 2. Let $W_t$ be the bilateral variation margin posted in cash earning interest at the risk-free rate. Write $W_t = W_t^+ - W_t^-$, $W_t^+$ the cash amount posted by party C to B, and $W_t^-$ B to C. Following similar derivation in Section 2 or Lou (2015), one can easily arrive at the following,

$$(\frac{\partial}{\partial t} + A_t)V - r_e(V - W) - rW - s_I L_I = 0$$
$$r_e(t) = r_b(t)I(V(t) - W(t) < 0) + r_c(t)I(V(t) - W(t) \geq 0).$$

For BCBS-IOSCO covered trades, $W(t)=V(t)$ so that the PDE reduces to

$$(\frac{\partial}{\partial t} + A_t)V^b - rV^b - s_I L_I^b = 0$$

where we have denoted superscript '$b$' as party B on the bid side, i.e., $V^b$ is its bid for the derivative. Note that the PDE is same as the Black-Scholes equation except for the IM funding cost term and that the same risk-free discount curve applies. For a call option, for instance, this would intuitively produce a lower fair price $V^b$ than the risk-free price $V^*$ without IM impact.

Party C is on the other side and the PDE for its fair price $V^c$ would look like, assuming the same IM funding spread,

$$(\frac{\partial}{\partial t} + A_t)V^c - rV^c + s_I L_I^c = 0$$

We then have trivially $V^b < V^* < V^c$, indicating that party B and C would never see the same prices.

Because of bilateral variation margin (and the added credit risk mitigation with initial margin), the industry standard OIS discount applies to BCBS-IOSCO trades. Now suppose two covered counterparties B and D attempt to price an OTC derivative in an attempt to recoup their respective IM funding costs. On a standalone basis, party B – the bidding party -- wants to deduct its funding cost from the bid price while on the ask side, party C wants to recover its funding cost by asking for a higher price. Neither side can effectuate a funding cost transfer, unless one side backs down and lets the other side act in a market making role. $V^c$ can then be taken as its ask price while $V^b$ bid price. For accounting purposes, the fair value could remain the risk-free price, leaving MVA or IM funding charges into bid/ask, should a firm choose to do so.

This has to be the case when the parties trade and clear through a CCP. CCP cleared trades are subject to the same PDE, bid or ask side, from the market maker's point of view. The only difference is that IM is now posted to the CCP rather than a segregated account with the counterparty. The CCP as the valuation agent can't be aware of either party's IM funding cost and the only agreeable valuation is to have it stripped from CCP valuation, resulting in the risk-free price.

We conclude that in a standalone trade, financial counterparties will not be able to transfer its funding cost to the other party, unlike when they trade with an uncollateralized client. Same is true for the whole OTC derivatives portfolio between financial counterparties.

**4.1. ISDA SIMM Connectivity**

For derivatives on a single asset risk factor (stock price or commodity or fx rate), the delta approximation formalized by ISDA can be built into the PDE. Take for example, the equity risk's delta initial margin is defined as the product of a bucketed risk weight and the derivatives' sensitivity (ISDA 2016), $L_I = S \left| \partial V / \partial S \right| RW_\%$. For equity indices, ISDA proposes a risk weight of 15. If we take SPX's volatility at 15%, the multiplier needed to calibrate $L_I$ to SIMM is 2.2.

Similarly, SIMM's curvature (or gamma) margin $L_{gamma}$ and vega margin $L_{vega}$ can be written as

$$L_{gamma} = (\frac{1}{2} \sigma^2 S^2 \partial^2 V / \partial S^2) \frac{RW_\%}{\sigma} R_{gamma}$$

$$L_{vega} = (\frac{1}{2} \sigma^2 S^2 \partial^2 V / \partial S^2) \frac{RW_\%}{\sigma} R_{vega}(T-t)$$

where R$_{gamma}$ is 0.5586, R$_{vega}$ 0.9218[10], RW$_\%$ is the SIMM's risk weight divided by 100.

---

[10] SIMM's IM for equity risk's vega boils down to $\mathrm{VR} = \frac{RW}{\phi^{-1}(0.99)} \sqrt{365/14} * (0.01 \frac{\partial V}{\partial \sigma})$

where $0.01 \frac{\partial V}{\partial \sigma}$ is vega sensitivity defined as price change over 1% change in volatility. The Black-Scholes formulae gives $\frac{\partial V}{\partial \sigma} = T \sigma S^2 \frac{\partial^2 V}{\partial \sigma^2}$, and IM_vega = VRW*VR where VRW for equity is 0.21. So

$\mathrm{IM\_vega} = 0.01 \sqrt{365/14} * 0.21 \frac{RW}{\phi^{-1}(0.99)} T \sigma S^2 \frac{\partial^2 V}{\partial \sigma^2} = (0.009218 \frac{RW}{\sigma} T) * \frac{1}{2} (\sigma S)^2 \frac{\partial^2 V}{\partial \sigma^2}$.

Putting these IM into the PDE leads to

$$\frac{\partial V}{\partial t} + S\frac{\partial V}{\partial S}(r - s_I RW_\% \operatorname{sgn}(\frac{\partial V}{\partial S})) + \tfrac{1}{2}\sigma^2 S^2 \frac{\partial^2 V}{\partial S^2}(1 - s_I(R_{gamma} + R_{vega}(T-t))\frac{RW_\%}{\sigma}) - rV = 0$$

Now the SIMM compliant BCBS_IOSCO equity derivatives can be fully priced in the PDE approach.

The multiplier does not enter here as the above is derived directly from SIMM's single asset model specifications. It however can be re-introduced to the delta term and calibrated, if the IM model has other risk factors, e.g., correlation. This would constitute a second approximation, as this number is different from the multiplier in Section 3 where it is calibrated to capture the IM difference between the risk neutral world and the real world, while here IM's gamma and vega components are separately captured.

For a portfolio of equity derivatives, adding or removing a position to the existing portfolio will incur netting and diversification effect under SIMM's two-level correlation structure. The incremental portfolio margin can be allocated to the position, e.g. via a net to gross proportional scheme. A multiplier can then be plugged in to reflect the allocated IM being smaller than the standalone IM.

**4.2. Solving PDE with IM funding cost**

The PDE capturing the initial margin funding cost under delta approximation can be solved in a tree or lattice model when the dimension is low such is the case at a single trade level or a granular portfolio, e.g., a commodity portfolio (of single underlying or rate) borrowing from CME's terminology. For plain vanilla swaps, payer or receiver, the Crank-Nicholson FD scheme (Lou 2016) can be used directly with an adjusted drift coefficient. For swap portfolios, an iteration is developed to cope with the discount rate switch and the delta switch of the initial margin funding cost. Computational results from a finite difference scheme will be shown in next section.

For other cases, the regression/simulation procedure (Lou 2016) can be modified to compute MVA alongside of CVA and FVA. The simulation procedure could accommodate IM calculation either endogenously under the risk-neutral measure (with a multiplier) or exogenously under the physical measure. With the former, we can reuse the regression equation intended for the switching rate to get the fair value shock given the shocks to the underlying risk factors; with the latter, exogenously provided historical shocks can be applied through a regression/compression scheme shown in Green and Kenyon (2015). A separate regression equation might be needed as the regression for the rate switch is of lower accuracy requirement as it cares only about the sign of value, while the regression for IM needs to extend a wide range in good accuracy.

For pricing a standalone new trade, the 99% shock can be applied in both directions. The worse loss of fair value is the estimated IM. For incremental pricing, as a firm could benefit from reduced IM and its funding cost, the multiplier can be made

---

As for the curvature part, CVR~0.5*14/365/T*VR. Plug in VR formula for the above. T cancels out and CVR has the $\frac{1}{2}(\sigma S)^2 \frac{\partial^2 V}{\partial \sigma^2}$ term. IM_gamma = (lamda+1)*CVR where lamda= ($\phi^{-1}$(0.99))^2-1 for single asset risk. Combining terms together leads to the IM_gamma coefficient.

algebraic with a positive value indicating the existing portfolio is long delta, negative short delta. A local estimation of IM could then check on its local delta's sign. If it's of the same sign, then IM would be increased and a funding cost needs to be built in. If it is of opposite signs, a funding benefit results.

## 5. Numerical Results

The sample results presented below are based on two one-factor Markovian short rate models, mixed-normal-lognormal model and Black-Karasinski model. The former has the form $dr_t = a(\theta - r)dt + \sigma(r)dW$, where the volatility function is defined as $\sigma(r) = \frac{r}{0.015}\sigma_2$, for r<0.015, $\sigma(r) = \sigma_2$ for $0.015 \leq r < 0.06$, and $\sigma(r) = \frac{r}{0.06}\sigma_2$ for r≥0.06, with parameters estimation on historical data $\sigma_2$=1.05%, a=0.05, $\theta$=0.044[11] (Hull et al 2014). Black-Karasinski model has $dx_t = \kappa(\mu - x)dt + \sigma dW$, $r_t = \exp(x_t)$, where $\kappa$, $\mu$, $\sigma$ are positive constants.

The risk-free rate is a constant spread below the LIBOR short rate, using a recent average 3 month LIBOR-OIS spread of 13 bp. Fix the long term average short rate at 4.4%, each model has three parameters to calibrate. We choose the three month LIBOR rate, 10 year ATM swap rate of 235.87 bp, and 10 year ATM cap at a yield value of 86.83 bp. Party B and C's cash funding curves are assumed to be a deterministic spread above the LIBOR short rate.

All numerical results are obtained by solving the PDE with Crank-Nicholson finite difference scheme.

### 5.1. Standalone uncollateralized trade pricing

To demonstrate, we set Party B's CDS (zero recovery rate) short rate to 75 bp and its funding basis 50 bp on top of LIBOR, approximated single 'A' rated. C's spread to LIBOR varies from 37.5 to 1000 bp, roughly reflecting credit rating range of "AAA/AA+" to "B". C's funding basis are 15, 30, 50 bp for the first three 'rating's and 80 bp for the rest. The MPR is 10 days and multiplier is set to 3 so that current delta approximation of the swap is approximately same as CME's initial margin amount.

Table 1 shows the swap npv and its adjustments (yield values in bp) with respect to the risk free price, broken down into CVA, DVA, CFA, DFA and finally MVA. Here we set $s_N$ to 50 bp (30 bp cost of fund + 20 bp for 10 year LIBOR swap and OIS swap spread.) For a "BBB" client, MVA is about 2 bp. MVA is higher for better rated clients, for the fixed funding cost of initial margin has to be carried longer time on average because there is less likelihood of early termination due to counterparty C's default. The total counterparty credit and funding adjustment (sum of CVA and CFA) obviously increases as counterparty's credit quality deteriorates.

---

[11] This number itself is a historical mean, used here without market price of risk adjustment as mean in the pricing measure.

Table 1. 10y ATM payer swap valuation adjustments with a single 'A' rated dealer facing a counterparty of hypothetic 'AAA/AA+' to 'B' rating.

| C-Libor | "Rating" | NPV | CVA | DVA | CFA | DFA | MVA |
|---|---|---|---|---|---|---|---|
| 37.5 | AAA/AA+ | -3.55 | 1.48 | 0.31 | 0.61 | 0.17 | 2.37 |
| 75 | AA/AA- | -4.98 | 2.39 | 0.32 | 1.19 | 0.18 | 2.32 |
| 125 | A | -6.83 | 3.59 | 0.33 | 1.91 | 0.18 | 2.26 |
| 250 | BBB | -11.12 | 7.18 | 0.35 | 2.8 | 0.2 | 2.12 |
| 500 | BB | -18.57 | 15.45 | 0.41 | 2.32 | 0.24 | 1.88 |
| 1000 | B | -30.05 | 28.15 | 0.56 | 1.65 | 0.32 | 1.55 |

For a receiver swap, MVA is about the same magnitude as the payer swap, e.g. 2.24 bp for a "BBB" client vs 2.12 bp for a payer swap shown above. For a single 'B' client, the difference is more pronounced, with receiver swap MVA at 2.07 bp vs payer's 1.55 bp. This can be explained as the exposure contingent on client's default is smaller in a receiver so the benefit of early termination is less than the payer.

MVA is almost linearly dependent on the multiplier. If the multiplier for instance is set at 4 to roughly match LCH.ClearNet's initial margin for the same 10 year ATM swap, the payer swap's MVA becomes 2.82 bp for a 'BBB' client, 0.70 bp on top of CME's MVA charge of 2.12 bp. In fact with multiplier at 1, the payer's MVA is 0.71 bp.

MVA's sensitivity to funding rate is also close to linear for the swap priced. With multiplier set at 1, a doubled initial margin funding rate (at 100 bp) results in payer's MVA at 1.41 bp, and 0.35 bp if the funding rate is reduced by half.

**5.2. Uncollateralized swap portfolio valuation**

The model developed is not limited to a single trade application. For a portfolio of swaps, the fair value V is the portfolio npv and all valuation adjustments (CVA, FVA, and MVA) computed are at the portfolio level, fully reflecting its netting effect. Table 2 shows a simple swap portfolio of a 5 year payer swap and a 10 year receiver swap (a hypothetic curve trade) with an uncollateralized party C of 80 bp funding basis and 295 bp of CDS spread. The curve trade's npv is -2.84 bp in effective yield while the sum of individually priced uncollateralized swaps is -5.36, a difference of 2.52 bp. Given B's funding basis is 50 bp, and CDS spread is 75 bp, CVA exhibits the largest netting effect (last row in Table 2.) of 1.95 bp, down from 2.84 bp to 0.89 bp after netting. With IM funding charge of 50 bp with multiplier of 3, MVA is reduced from 2.66 bp to 1.74 bp.

Table 2. XVA netting effects shown for a 5 and 10y curve trade.

| | NPV | CVA | DVA | CFA | DFA | MVA | TVA |
|---|---|---|---|---|---|---|---|
| 5y payer | -19.74 | 1.2 | 0.31 | 0.29 | 0.18 | 0.47 | 1.47 |
| 10y rec | 14.38 | 1.64 | 3.09 | 0.41 | 1.65 | 2.19 | -0.5 |
| Sum | -5.36 | 2.84 | 3.4 | 0.7 | 1.83 | 2.66 | 0.97 |
| Portf | -2.84 | 0.89 | 2.86 | 0.21 | 1.52 | 1.74 | -1.54 |
| difference | 2.52 | -1.95 | -0.54 | -0.49 | -0.31 | -0.92 | -2.51 |

Last column shows total valuation adjustment (TVA=CVA-DVA+CFA-DFA+MVA.)

Table 3 shows another strategy, a combination of long 7 year cap with strike at 2.95% and short a floor at 2.36%, which creates shifted positive and negative payoff. For the long cap when priced standalone, there is no DVA and DFA, and for the short floor, no CVA and CFA. MVA however is asymmetric and exists with both positive exposure (long cap) and negative exposure (short floor.)

Table 3. XVA netting effects for a long cap short floor combination.

|  | NPV | CVA | DVA | CFA | DFA | MVA | TVA |
|---|---|---|---|---|---|---|---|
| Cap | 32.64 | 6.43 | 0 | 1.49 | 0 | 0.669 | 8.589 |
| Floor | -38.84 | 0 | 0.76 | 0 | 0.42 | 0.2493 | -0.9307 |
| Sum | -6.2 | 6.43 | 0.76 | 1.49 | 0.42 | 0.9183 | 7.6583 |
| Portf | -4.65 | 4.41 | 0.17 | 1.02 | 0.1 | 0.9222 | 6.0822 |
| difference | 1.55 | -2.02 | -0.59 | -0.47 | -0.32 | 0.0039 | -1.5761 |

Table 4 compares FD with Monte Carlo simulation for a 10 year receiver swap. The difference between these two solutions are less than 0.03 bp for swap npv yield value and 0.02 to 0.07 bp for MVA.

Table 4. Comparison of FD and Monte Carlo regression/simulation (MC-RS) solution.

| C-Libor | "Rating" | FD-NPV | FD-MVA | MC-RS NPV | MC-RS MVA |
|---|---|---|---|---|---|
| 37.5 | AAA/AA+ | 2.5861 | 2.307 | 2.566 | 2.333 |
| 75 | AA/AA- | 2.4565 | 2.295 | 2.435 | 2.323 |
| 125 | A | 2.2872 | 2.28 | 2.265 | 2.311 |
| 250 | BBB | 1.881 | 2.243 | 1.856 | 2.28 |
| 500 | BB | 1.1361 | 2.176 | 1.109 | 2.224 |
| 1000 | B | -0.1285 | 2.066 | -0.155 | 2.134 |

Following Green and Kenyon (2015)'s construction of a large test swap portfolio to compare FVA and MVA's magnitude, Table 5 shows MVA and CRA when the portfolio is 90%, 50%, and 10% payer, between two BCBS-IOSCO counterparties of the same credit spread of 125 bp. In the case of a predominantly payer swap portfolio of net notional of 765.7 million, the npv is -872.8 bp of the gross (not annualized). Unannualized CRA is -28.6 bp vs MVA of 25 bp. The sum of DVA and CFA is roughly about twice as much as MVA, similar to Green and Kenyon. This type of comparison, however, obviously depends on what IM funding spread is used, as shown in the next example.

Table 5. Comparison of CRA and MVA (with multiplier of 3) for three portfolios of 1000 swaps with tenors from 3 month to 30 years, gross notional of 1,002.1 million.

| Payer% | Net Ntl (mm) | NPV (bp) | CVA | DVA | CFA | DFA | CRA | MVA |
|---|---|---|---|---|---|---|---|---|
| 90% | 765.7 | -872.8 | 18.8 | 37.1 | 9.6 | 19.9 | -28.6 | 25 |
| 50% | -17.8 | 14.3 | 1.8 | 0.3 | 0.9 | 0.1 | 2.3 | 0.6 |
| 10% | -770.7 | 805 | 35.4 | 18.4 | 19.2 | 9.5 | 26.7 | 18.8 |

**5.3. ISDA SIMM pricing for equity risk**

For equity, commodity, and fx risk factors, incorporating IM and its funding cost in the PDE is useful, allowing existing models to be reused with little modification. Table 6 shows MVA for at-the-money European call options of 1 year and 2 year expiry, under some typical IM funding scenarios. In the first scenario, we assume 50% of the IM is funded at secured rate of 0.5% and the rest at the firm's senior unsecured rate of 1%. The one year call's MVA is only 15 cent, out of its risk free fair value of 20.1346. The other four scenarios range from full leverage ('0% Lev' row), BASEL III's minimum 3% required leverage ratio, 6% for systemically important financial institutions (SIFIs), and no leverage, i.e., full equity funding at a return on equity (ROE) target of 15%. Obviously as uses of leverage decrease from full leverage to no leverage, the effective IM funding spread increases from 1% to 15%. MVA increases accordingly from 20 cents to 2 dollars and 92 cents.

Under ISDA SIMM, MVA breaks down into delta and vega/gamma contributions, the latter shown in column 'MVA-gv', which is approximately 25% of the total MVA under all funding scenarios considered. Applying an extra multiplier of 3, for instance, increases MVA almost by three times. MVA doubles when the expiry is extended to 2 years. The last two columns then show the bid and ask prices.

Column 'MVA-M 0.234' applies a multiplier of 0.234, calculated as the proportion of the incremental IM of adding a new stock option to an existing portfolio of 100 stocks randomly scattered in SIMM's 12 buckets, to the standalone IM. The incremental MVA is much smaller than the standalone MVA shown in column 'MVA-dgv'.

Table 6. MVA of European call options, $S=K=100$, $vol=50\%$, $r=1\%$, $T=1$ or 2 years. 'MVA-dgv' column shows one year option's MVA inclusive of delta, curvature gamma and vega risks; 'MVA-gv' gamma and vega risk only. 'MVA -M3' applies a multiplier of 3 to all risks; and 'MVA 2y dgv' shows 2 year option's all-inclusive MVA.

| IM funding | Sprd (%) | MVA-dgv | MVA-gv | MVA-M0.234 | MVA 2y dgv | Bid | Ask |
|---|---|---|---|---|---|---|---|
| 50% Sec-0% Lev | 0.75 | 0.15 | 0.04 | 0.04 | 0.32 | 19.98 | 20.21 |
| 0% Lev | 1 | 0.20 | 0.05 | 0.09 | 0.42 | 19.77 | 20.32 |
| 3% Lev | 1.42 | 0.28 | 0.07 | 0.07 | 0.60 | 19.85 | 20.28 |
| 6% Lev | 1.84 | 0.37 | 0.09 | 0.05 | 0.77 | 19.93 | 20.24 |
| 100% Lev | 15 | 2.92 | 0.75 | 0.70 | 6.00 | 17.22 | 21.74 |

## 5.4. Incremental MVA and CME-LCH basis

The IM multiplier introduced can be used to calibrate to a CCP's IM calculation on a given commodity portfolio. Let $\eta_p$ denote the calibrated portfolio multiplier. Now suppose a new trade is to be added to the existing portfolio. An incremental pricing scheme could be to apply $\eta_p$ to the new trade as if it is a standalone trade solved by its standard pricing model, e.g., a finite difference model. Here we attempt to explain the CME-LCH basis spread with the multiplier's variability.

Suppose a dealer hedges a CME swap with an asset manager with another b/d in LCH. The dealer receives the CME-LCH basis spread to cover its cost of funding its dual IM posting at both CME and LCH. Obviously the basis is proportional to the allocated multiplier $\eta_p$. When the pace of new trade flow is slow, $\eta_p$ is small, resulting in a small basis. But when the flow increases multiple times, the allocated $\eta_p$ is much greater, leading to multiplication of the basis, providing support for a popular explanation of recent widening of CME-LCH swap basis, e.g. 30 year swap spread close to 3.7 bp on November 17, 2015, which is related to a jump of one-sided flow of asset managers swapping out fixed rates to b/d in CME and b/d's back-to-back hedging in LCH (Khwaja, 2015).

Figure 1 shows a 10 year ATM swap's MVA (in yield value, bp) on the b/d receiving fixed with CME and MVA due to paying fixed with LCH. The IM funding cost is fixed at 100 bp, assuming the b/d can fund half of the IM at a secured cost of 40 bp, and the remaining half split between its equity capital with 10% ROE and unsecured debt rate of 125 bp modulated at 25 leverage ratio. Both MVAs increase as $\eta_p$ goes up, assuming the same $\eta_p$ in both CME and LCH. Adding up these two MVAs gets the corresponding swap rate basis spread between CME and LCH. The left end has $\eta_p=0.1088$ and a basis spread of 0.22 bp, a level prior to the widening. The peak CME-LCH 10y basis of 2.887 bp observed in November 2015 corresponds to $\eta_p=1.4142$.

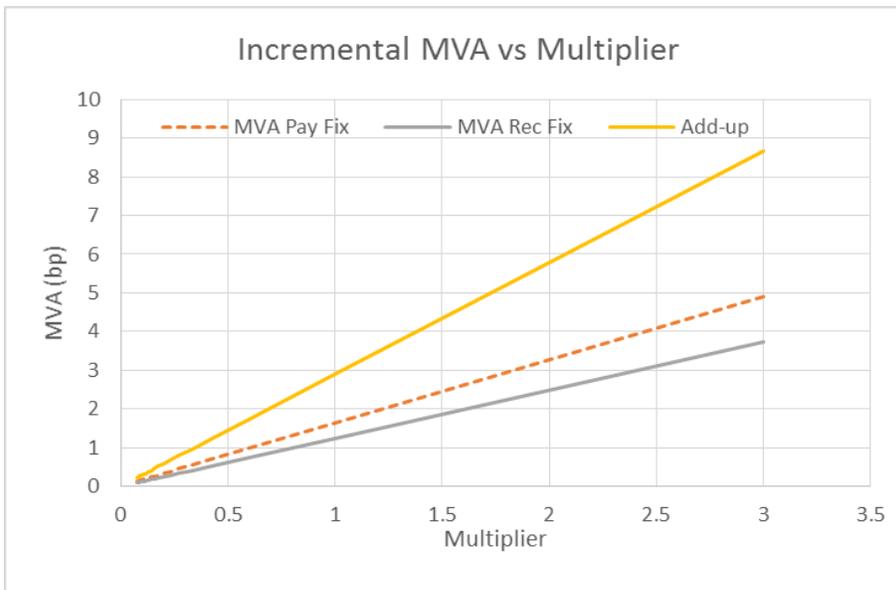

Figure 1. Inter-CCP MVA (MVA on paying fixed with one CCP and MVA on receiving fixed with another) varies on multiplier.

# 6. Conclusion

By adopting an idealized redundant initial margin account, the liability-side pricing theory is extended to cope with IM funding costs. The combination of computing IM by a delta-approximated VaR in the pricing measure and a multiplier calibrated to external IM models on real portfolios allows a balance of meeting CCP or regulatory requirements and having numerically efficient implementations. For uncollateralized customer trades, margin valuation adjustment (MVA) is defined as the liability-side discounted funding costs on the expected margin profile when it can be exogenously determined, or a component of an all-in price governed by an extended PDE and solved by finite difference methods. At 10-day 99-percentile, and 150 bp margin funding cost, a standalone at-the-money uncollateralized client swap of 10 year maturity shows about 2 basis point equivalent charge to be fully transferred to the client.

For CCP cleared or BCBS-IOSCO covered non-centrally cleared OTC derivatives, IM funding charge is effectively a bi-way tax, unless one party serves as the market maker whereas the funding charge is transferred in the form of a bid-ask spread, which can be solved from the PDE with links to external IM models such as ISDA SIMM. This PDE approach is particularly fitting for equity, commodity, or fx derivatives portfolio, with SIMM's vega and gamma contribution to initial margin captured in the PDE's gamma term.

The multiplier can also be used for portfolio level incremental pricing, as used to illustrate recent CME-LCH basis spread widening due to jumps in MVA following dealers' hedging of customer flows.

**References**

BCBS-IOSCO, 2015, Margin requirements for non-centrally cleared derivatives, March 2015 final document, Basel Committee on Banking Supervision, Board of the International Organization of Securities Commissions., www.bis.org/bcbs/publ/d317.htm.

Brigo, D. and A. Pallavicini, 2014, CCP cleared or bilateral CSA trades with initial margin/variation margins under credit, funding and wrong-way risks: a unified valuation approach, ssrn working paper.

Burgard, C. and M. Kjaer, 2011, Partial differential equation representations of derivatives with bilateral counterparty risk and funding costs, J of Credit Risk, **7**(3), pp.75-93.

Green, A. and C Kenyon, 2015, MVA by replication and regression, *Risk*, May, pp 82-87.

Hull, J., A. Sokol and A. White, 2014, Short Rate Joint Measure Models, *Risk* October, pp. 59-63.

ISDA, 2016, ISDA SIMM[TM.1] Methodology, version 3.15.

Khwaja, Amir, 2015, CME-LCH Basis Spread, https: //www.clarusft.com/cme-lch-basis-spreads, Clarus Financial Technology Newsletter, May 20.

Lou, Wujiang, 2015, CVA and FVA with Liability-Side Pricing, *Risk*, August, pp.48-53.

Lou, Wujiang, 2016, Liability-side Pricing of Swaps, *Risk*, April, pp. 61-66.

Stein, H., 2015, Two Measures for the Price of One, *Risk*, March.